\documentclass[openacc]{rstransa}

\DeclareRobustCommand\openone{\leavevmode\hbox{\small1\normalsize\kern-.33em1}}

\usepackage{bm}
\usepackage{enumerate}
\usepackage{color}
\usepackage{mathptmx}      

\begin{document}

\title{Quantum theory as plausible reasoning applied to data obtained by robust experiments}

\author{
H. De Raedt$^{1}$,  M.I. Katsnelson$^{2}$, and K. Michielsen$^{3,4}$}

\address{$^{1}$Zernike Institute for Advanced Materials, \\ University of Groningen, \\
NL-9747 AG Groningen, The Netherlands
\\
$^{2}$Institute of Molecules and Materials, \\Radboud University,\\
NL-6525 ED Nijmegen, The Netherlands
\\
$^{3}$Institute for Advanced Simulation, J\"ulich Supercomputing Centre,
Forschungszentrum J\"ulich, D-52425 J\"ulich, Germany
\\
$^{4}$RWTH Aachen University, D-52056 Aachen, Germany}

\subject{xxxxx, xxxxx, xxxx}


\corres{H. De Raedt\\
\email{h.a.de.raedt@rug.nl}}

\begin{abstract}
We review recent work that employs the framework of logical inference to establish
a bridge between data gathered through experiments and their objective description in terms of human-made concepts.
It is shown that logical inference applied to experiments for which the observed events are independent
and for which the frequency distribution of these events is robust with respect to small changes of
the conditions under which experiments are carried out yields,
without introducing any concept of quantum theory,
the quantum theoretical description in terms of the Schr\"odinger or the Pauli equation,
the Stern-Gerlach or Einstein-Podolsky-Rosen-Bohm experiments.
The extraordinary descriptive power of quantum theory then follows from the fact that it is plausible reasoning,
that is common sense, applied to reproducible and robust experimental data.
\end{abstract}



\begin{fmtext}


\end{fmtext}
\maketitle

\section{Introduction}\label{sec1}

Quantum theory is unsurpassed as a description of the data produced by many very different experiments in (sub)-atomic,
molecular and condensed matter physics,  quantum optics etc.
A large body of work focuses on different interpretations of the quantum formalism~\cite{BOHM52,PENA96,BALL70,HOME97,BALL01,CAVE07,BUB07,FUCH13}
and its derivations from different sets of axioms~\cite{KHRE05,HARD07,KHRE09,KHRE10,CHIR10,CHIR11,BRUk11,MASA11,ORES12,KLEI12a,KAPU13,PENA15}
but offers no explanation of the success of quantum theory that goes beyond ``that is because of the way it is''.
In the present paper, we review recent work that de-mystifies the extraordinary power of quantum theory~\cite{RAED13b,RAED14b,RAED15b}
by formalizing the human thought processes by which relations
between data gathered through experiments and objective descriptions in terms of human-made concepts may be established.
A basic premise of this approach is that scientific theories are the result of cognitive processing
of the discrete events which are registered by our sensory system,
of the logical and/or cause-and-effect relations between those events,
and the use of metaphors to make abstractions and construct concepts.
This form of cognitive processing may be expressed in terms of the algebra of logical inference (LI),
a mathematical framework that facilitates rational reasoning when there is uncertainty~\cite{COX46,COX61,TRIB69,SMIT89,JAYN03}.
Statistical mechanics can be given an information-theoretical justification by viewing the former as a problem of LI,
thereby establishing a relation between the information-theoretic entropy~\cite{SHAN49}
and the thermodynamics entropy~\cite{JAYN57a,JAYN57b}.
The resulting maximum-entropy principle~\cite{JAYN57a,JAYN57b,TRIB69,JAYN03} has recently been generalized to a principle
of entropic dynamics, a framework in which dynamical laws can be derived~\cite{CATI14,CATI15}.

The LI approach, being extended logic, is not bound by ``laws of physics'' and does not require
assumptions such as ``the observed events are signatures of an underlying objective reality
-- which may or may not be mathematical in nature'', ``all things physical are information-theoretic in origin'',
``the universe is participatory'', etc.
It yields results that are unambiguous and independent of the individual subjective judgment,
providing a rational explanation for the extraordinary descriptive power of quantum theory
and it also provides strong support for Bohr's statement~\cite{BOHR99}
``The physical content of quantum mechanics is exhausted by its power to formulate
statistical laws governing observations under conditions specified in plain language''.

LI applies to situations where there may or may not be causal relations between the events~\cite{TRIB69,JAYN03}.
Extracting cause-and-effect relationships from empirical evidence is a highly non-trivial problem.
In general, LI does not establish cause-and-effect relationships~\cite{JAYN03,PEAR09},
although rational reasoning about these relations should comply with the rules of LI.
Furthermore, a derivation of a quantum theoretical description from LI principles
does not prohibit the construction of cause-and-effect mechanisms that create the {\it impression}
that these mechanisms produce data that can be described by quantum theory~\cite{THOO07,RAED05b}.
In fact, there is a substantial body of work demonstrating that it is indeed possible to construct simulation models
which reproduce, on an event-by-event basis, the results of interference/entanglement/uncertainty
experiments with single photons/neutrons~\cite{MICH11a,RAED12a,RAED12b,RAED14a,MICH14a}.
This demonstration does not imply the reality of hidden variables or something like that.

The LI approach which we review here leads to the view that quantum theory is a phenomenological theory
which can be derived from a set of simple general principles, not axioms,
in a way that is independent of any (strictly speaking, unknown)
``more microscopic'' level of description.
Therefore its power does not depend
on whether there exists an underlying classical world with some hidden variables, or not.
In this sense, there is a clear parallel with Einstein's view on thermodynamics.
Einstein did not regard thermodynamics as a constructive theory, an attempt to build a picture of complex phenomena
out of some relatively simply propositions but rather as a theory of principles based on empirically
observed properties of phenomena, independent of a particular underlying model~\cite{KLEI67}.

The paper is structured as follows.
Section~\ref{sec2} briefly recapitulates the basic elements of LI
and reviews applications to the Stern-Gerlach (SG) and Einstein-Podolsky-Rosen-Bohm (EPRB)
experiment and a particle in a potential.
It shows how the LI approach directly leads to the probabilities for observing the events
without invoking any concept of quantum theory.
In Section~\ref{sec3}, we discuss two methods for transforming the solutions
obtained through LI into the equations that we know from quantum theory.
A summary and discussion of more general aspects of the work
presented in this paper are given in Section~\ref{sec4}.

\section{Logical inference}\label{sec2}

The key concept of the LI approach is the plausibility, denoted by $P(A|B)$, which
in general, expresses the degree of believe of an individual that
proposition $A$ is true, given that proposition $B$ is true~\cite{POLY54,COX61,TRIB69,JAYN03}.
The plausibility $P(A|B)$ is an intermediate mental construct
that serves to carry out inductive logic, that is rational reasoning,
in a mathematically well-defined manner~\cite{TRIB69,JAYN03}.

The algebra of logical inference can be derived from three so-called ``desiderata'', namely
(i) {\it plausibilities are represented by real numbers},
(ii) {\it plausibilities must exhibit agreement with rationality}, and
(iii) {\it all rules relating plausibilities must be consistent}~\cite{COX61,TRIB69,SMIT89,JAYN03}.
These three desiderata only describe the essential features of the plausibilities
and are not a set of axioms which plausibilities have to satisfy.
It is a most remarkable fact that these three desiderata suffice to uniquely determine
the set of rules by which plausibilities may be manipulated~\cite{COX61,TRIB69,SMIT89,JAYN03}.
It can be shown~\cite{COX61,TRIB69,SMIT89,JAYN03} that plausibilities may be chosen
to take numerical values in the range $[0,1]$ and these values are related by three rules, namely
(1) $P(A|Z)+P({\bar A}|Z)=1$ where ${\bar A}$ denotes the negation of proposition $A$
and $Z$ is a proposition assumed to be true,
(2) the ``product rule'' $P(AB|Z)=P(A|BZ)P(B|Z)=P(B|AZ)P(A|Z)$ where
the ``product'' $BZ$ denotes the logical product (conjunction)
of the propositions $B$ and $Z$, and
(3) $P(A{\bar A}|Z)=0$ and $P(A+{\bar A}|Z)=1$
where the ``sum'' $A+B$ denotes the logical sum (inclusive disjunction)
of the propositions $A$ and $B$~\cite{COX61,TRIB69,SMIT89,JAYN03}.
The algebra of logical inference, as defined by the rules (1)--(3),
contains Boolean algebra as a special case and
is the foundation for powerful tools such as the maximum entropy method
and Bayesian analysis~\cite{TRIB69,JAYN03}.
The rules (1)--(3) are unique~\cite{TRIB69,SMIT89,JAYN03}:
any other rule which applies to plausibilities represented by real numbers
and is in conflict with rules (1)--(3) will be at odds with rational reasoning and consistency,
as embodied by the desiderata (i)--(iii).
It should be mentioned here that it is not allowed
to define a plausibility for a proposition
conditional on the conjunction of mutual exclusive
propositions: reasoning on the basis of two or more contradictory premises is out of the scope of LI.

The applications of LI which we review in the present paper
describe phenomena in a manner which is independent of individual subjective judgment.
Therefore, to differentiate between the ``objective'' and ``subjective'' mode of application of LI,
we will refer to the plausibility as ``inference-probability'' or ``i-prob'' for short.
A more extensive discussion and arguments for distinguishing between plausibility,
inference-probability, and Kolmogorov probability can be found in Ref.~\cite{RAED14b}.
For the purpose of the present review, it is sufficient
if the reader does not think of the i-prob as a frequency or probability
in the traditional mathematical sense but merely as numerical measure for the fact
that proposition $A$ is true, given that proposition $B$ is true.

In real experiments there is always uncertainty about some
factors which may or may not influence the outcome of the measurements:
it is presumptuous to assume that we know ``everything'' about these factors.
In particular, if experiment shows that
(a) there is uncertainty about each individual event, and
(b) the conditions under which the experiment is carried out
are also subject to uncertainties,
then the data collected in such an experiment
cannot be described by the traditional theories of classical physics, the reason being
that the theoretical description of ``classical physics'' assumes
that there is absolute certainty about the outcome of each individual experiment on each individual object.
In contrast, the LI approach is well-suited to deal with uncertainties but, as will be explicitly shown later,
to render the resulting description free of individual subjective judgment,
it is necessary to assume that
(c) the frequencies with which events are observed are reproducible
and robust (to be discussed later) against small changes in the conditions.
Furthermore, the LI approach only yields a quantum theoretical
description if in addition we assume that
(d) individual events are independent, meaning that knowing any (necessarily finite)
set of events does not help to increase the certainty by which we can
predict another (past or future) event that does not belong to the set.

If the experimental data complies with requirements (a)--(d),
application of LI rather straightforwardly yields basic equations of quantum theory.
The LI derivation of these equations has a generic structure.
The first step is to list the features of the experiment that are deemed
to be relevant and to introduce the i-probs of the individual events.
The second step is to impose the condition that the experiment
yields robust, reproducible results, not on the level of individual events,
but on the level of the frequencies of observing many events
and, depending on the problem, to impose other constraints about e.g.
the fact that the particle moves etc.
As an example of such a constraint, we will use a natural requirement that the equations of classical mechanics
should be as accurate as possible for the {\it average} results of many quantum experiments.
The result of the second step is a functional of the i-prob.
The third step is to solve the robust optimization problem
defined in terms of this functional and, optionally, to transform the formal solution
in terms of i-probs into linear equations
which we recognize as basic equations of quantum theory.

\begin{figure}[t]
\centering\includegraphics[width=\hsize]{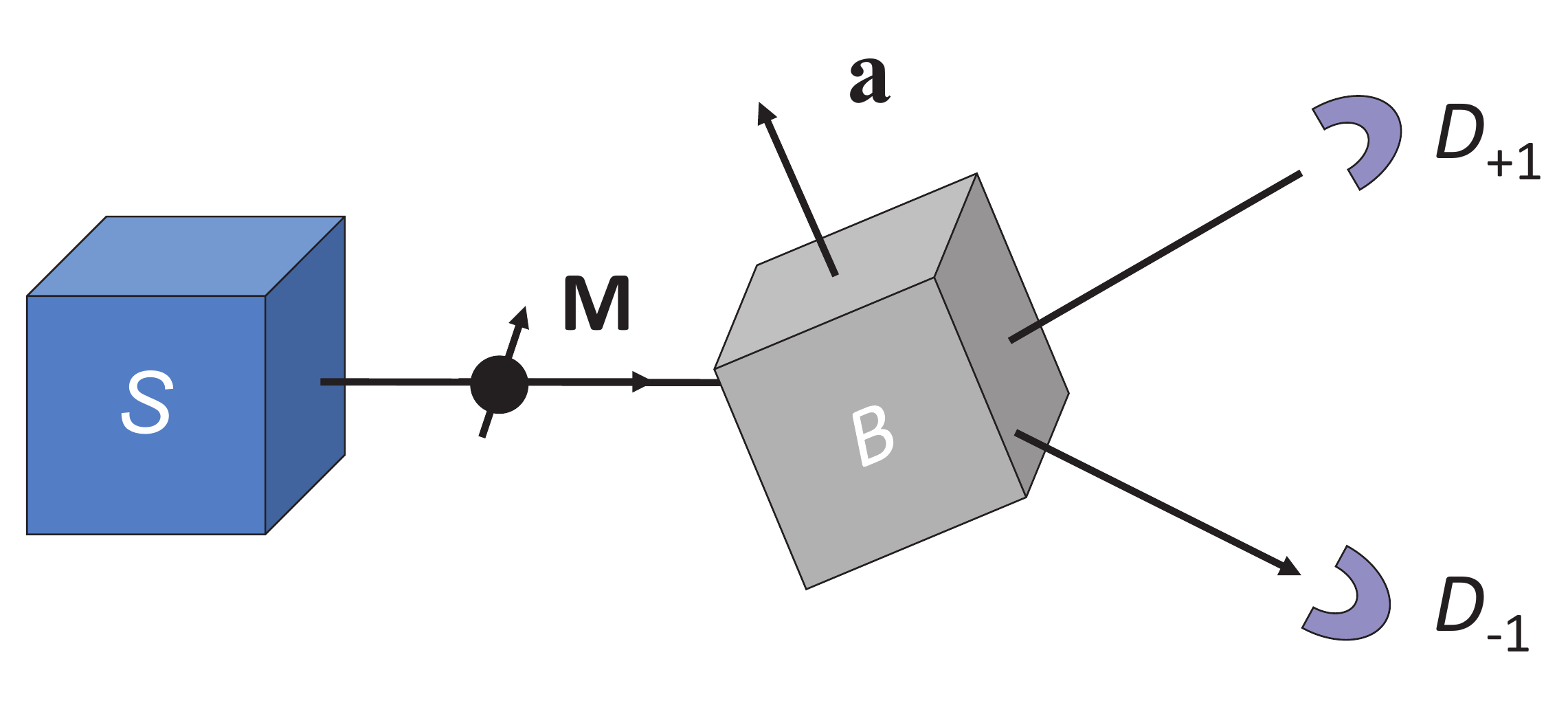}
\caption{(Color online)
Diagram of the SG experiment.
The source $S$, activated at times labeled by $i=1,2,\ldots,N$,
sends a particle carrying a magnetic moment $\mathbf{M}$
to the magnet $B$ with its (inhomogeneous) magnetic field oriented along the direction $\mathbf{a}$.
Depending on the relative directions of $\mathbf{a}$ and $\mathbf{M}$,
the particle is detected with 100\% certainty by either
$D_{+1}$ or $D_{-1}$.
}
\label{SGthought}
\end{figure}

\subsection{Application: Stern-Gerlach experiment}\label{sec2a}

The application of LI to event-based phenomena follows a particular pattern which
is best illustrated by considering the simplest case, namely the SG experiment of which the schematic is shown in Fig.~\ref{SGthought}.
In the SG experiment, there are two different outcomes which we label by the variable $x$ taking the values $x=+1$ or $x=-1$.
In a SG experimentthe source is activated at discrete times labeled by $i=1,\ldots,N$, resulting in
time series of detection events $x_i=\pm1$.
The first step in the LI treatment is to assign to an individual event $x=\pm1$,
an i-prob $P(x|\mathbf{a},\mathbf{M},Z)$ to observe that event.
Here, $\mathbf{a}$ and $\mathbf{M}$ are shorthands for the proposition
that (within a small range) the directions of the magnet and of the magnetic moment of the particle
are indeed $\mathbf{a}$ and $\mathbf{M}$, respectively,
and that the proposition $Z$, representing all the other conditions under which the experiment is performed, is true.
It is assumed that the conditions represented by $Z$ are fixed and identical for all experiments.

Assuming that the observed counts do not depend on the orientation of the chosen reference frame,
$P(x|\mathbf{a},\mathbf{M},Z)$ can only depend on $\mathbf{a}\cdot\mathbf{M}$
(by construction $\vert\mathbf{a}\vert=1$ and $\vert\mathbf{M}\vert=1$).
Hence, we must have $P(x|\mathbf{a},\mathbf{M},Z)=P(x|\mathbf{a}\cdot\mathbf{M},Z)=P(x|\theta,Z)$
where $\cos\theta=\mathbf{a}\cdot\mathbf{M}$.
This assumption is necessary to consider $\mathbf{M}$ as the direction of the magnetic moment of the particle
whereas only $\mathbf{a}$ is known from experiment.
Such symmetry requirements are very important for our construction as
they establish relations between what is measured (position of detector)
and what is supposed to be measured (characteristics of a particle).
It is expedient to write $P(x|\theta,Z)$ as
\begin{equation}
P(x|\theta,Z)=\frac{1+xE(\theta)}{2}
\quad,\quad
E(\theta)=\sum_{x=\pm1}xP(x|\theta,Z)
.
\label{sg1}
\end{equation}
According to assumption (d) there is no relation between the actual values
of $x_n$ and $x_{n'}$ if $n\not=n'$. With this assumption, repeated application of the product rule yields
\begin{eqnarray}
P(x_1,\ldots,x_N|\theta,Z)
&=&
\prod_{i=1}^{N}P(x_i|\theta,Z)
.
\label{sg2}
\end{eqnarray}
Repeating the experiment $N$ times yields $n_{x}$ events of the type $\{x\}$ ($n_{+1}+n_{-1}=N$)
and the i-prob to observe the compound event $\{n_{+1},n_{-1}\}$ is given by~\cite{RAED14b}
\begin{equation}
P(n_{+1},n_{-1}|\theta,N,Z)=N!\prod_{x=\pm1} \frac{P(x|\theta,Z)^{n_{x}}}{n_{x}!}
.
\label{robu0}
\end{equation}
Although the individual events may be expected to change from run to run, for sufficiently large $N$
the numbers $\{n_{+1},n_{-1}\}$ should exhibit some kind of robustness with respect to small changes of $\theta$.
Otherwise the $\{n_{+1},n_{-1}\}$ would vary erratically with $\theta$ and these ``irreproducible'' experiments would be discarded.
Obviously, the expected robustness with respect to small variations should be reflected in the expression
of the i-prob to observe the data (within the usual statistical fluctuations).

If the outcome of the experiment is indeed described by the i-prob Eq.~(\ref{robu0}) and the experiment is supposed
to yield reproducible, robust results, small changes of $\theta$ should not have a drastic effect on the outcome.
Let $H_0$ and $H_1$ be the hypothesis that the data $\{n_{+1},n_{-1}\}$ is observed if
the angle between the unit vector $\mathbf{a}$ is $\theta$ and $\theta+\epsilon$, respectively.
The evidence $\mathrm{Ev}$ of hypothesis $H_1$, relative to hypothesis $H_0$, is defined by~\cite{TRIB69,JAYN03}
\begin{equation}
\mathrm{Ev}=\ln\frac{P(n_{+1},n_{-1}|\theta+\epsilon,N,Z)}{P(n_{+1},n_{-1}|\theta,N,Z)}
,
\label{robu1}
\end{equation}
where the logarithm serves to facilitate the algebraic manipulations.
If $H_1$ is more (less) plausible than $H_0$ then $\mathrm{Ev}>0$ ($\mathrm{Ev}<0$).
The absolute value of the evidence, $|\mathrm{Ev}|$ is a measure for the robustness of the description
(the sign of $\mathrm{Ev}$ is arbitrary, hence irrelevant): the smaller $|\mathrm{Ev}|$
the more robust the experiment is for small changes of $\theta$.

The problem of determining the most robust description of the experimental data may now be formulated
as follows: search for the i-prob's $P(n_{+1},n_{-1}|\theta,N,Z)$
which minimize $|\mathrm{Ev}|$ for all possible $\epsilon$ ($\epsilon$ small) and for all possible $\theta$.
The clauses ``for all possible $\epsilon$ and  $\theta$'' render the
minimization problem an instance of a robust optimization problem.
The robust optimization problem has a trivial solution, namely
$P(n_{+1},n_{-1}|\theta,N,Z)=P(n_{+1},n_{-1}|N,Z)$, which
can only describe experiments for which $\{n_{+1},n_{-1}\}$ show no dependence on $\theta$.
Experiments which produce results that do not depend on the conditions
seem fairly pointless and therefore we explicitly exclude i-prob's
that are constant with respect to changes of the conditions.
It is not difficult to show~\cite{RAED14b} that our concept of a robust experiment implies that the i-prob's which
describe such experiment can be found by minimizing $|\mathrm{Ev}|$, subject to the constraints that
(C1) $\epsilon$ is small but arbitrary, (C2) not all i-prob's are independent of $\theta$, and
(C3) $|\mathrm{Ev}|$ is independent of $\theta$~\cite{RAED13b,RAED14b,RAED15b}.
Omitting terms of ${\cal O}(\epsilon^3)$, minimizing $|\mathrm{Ev}|$ while taking into account the
constraints (C2) and (C3) amounts to finding the i-prob's $P(x|\theta,Z)$ which minimize~\cite{RAED14b}
\begin{equation}
I_{F}= \sum_{x\pm1}
\frac{1}{P(x|\theta,Z)}
\left(\frac{\partial P(x|\theta,Z)}{\partial\theta}\right)^2
,
\label{robu6}
\end{equation}
subject to the constraint that $\partial P(x|\theta,Z)/\partial\theta\not=0$.
The r.h.s. of Eq.~(\ref{robu6}) is the Fisher information for the problem at hand
and because of the constraint (C3), should not depend on $\theta$.

\begin{center}
\medskip
\framebox{
\parbox[t]{0.95\hsize}{%
In the course of deriving Eq.~(\ref{robu6}), our criterion of robustness enforces
the intuitively obvious assignment $P(x|\theta,Z)=n_{x}/N$,
establishing the relation between the epistemological concept (i-prob) and
the physically measurable quantity (frequency of outcomes).
It is at this point that the possibility to view the i-prob as a ``subjective'' assignment is eliminated~\cite{RAED14b}.
}}
\medskip
\end{center}

Using Eq.~(\ref{sg1}) we can rewrite Eq.~(\ref{robu6}) as
$I_F=\big(\partial E(\theta)/\partial \theta\big)^2\big/\big((1-E^2(\theta)\big)$,
yielding $E(\theta)=\cos\left(\theta\sqrt{I_F}+\phi\right)$
where $\phi$ is an integration constant.
As $E(\theta)$ is a periodic function of $\theta$
we must have $\sqrt{I_F}=K$ where $K$ is an integer and hence
$E(\theta)=\cos\left(K\theta+\phi\right)$.
The solution $K=I_F=0$ is excluded from further consideration
because it describes an experiment in which the frequency distribution of the observed data
does not depend on $\theta$ (see constraint (C2)).
Therefore, the physically relevant, nontrivial solution with minimum Fisher information corresponds to $K=1$.
Furthermore, as $E(\theta)$ is a function of
$\mathbf{a}\cdot\mathbf{M}=\cos\theta$ only, we must have $\phi=0,\pi$.
Therefore, for the Stern-Gerlach experiment, the solution of the robust optimization problem reads
\begin{eqnarray}
P(x|\mathbf{a}\cdot\mathbf{M},Z)=
P(x|\theta,Z)=\frac{1\pm x\mathbf{a}\cdot\mathbf{M}}{2}
.
\label{sg5}
\end{eqnarray}
The $\pm$ sign in Eq.~(\ref{sg5}) reflects the fact that
the mapping between $x=\pm1$ and the two different directions
is only determined up to a sign.

Comparing Eq.~(\ref{sg5}) with the quantum theoretical expression (which is exactly the same)
demonstrates that Born's rule, one of the postulates
of quantum theory, appears as a consequence of LI applied to robust experiments~\cite{RAED14b,RAED15b}.
In the LI approach, Eq.~(\ref{sg5}) is not postulated
but follows from the assumption that the (thought) experiment that is being
performed yields the most reproducible results,
revealing the conditions for an experiment to produce data which
is described by quantum theory.

\subsection{Application: EPRB experiment}\label{sec2b}

The LI treatment of the EPRB experiment is, in essence,
the same as the one of the SG experiment.
Therefore, we only discuss the main assumptions and present the results.
Technical details can be found elsewhere~\cite{RAED14b}.

\begin{figure}[t]
\centering\includegraphics[width=\hsize]{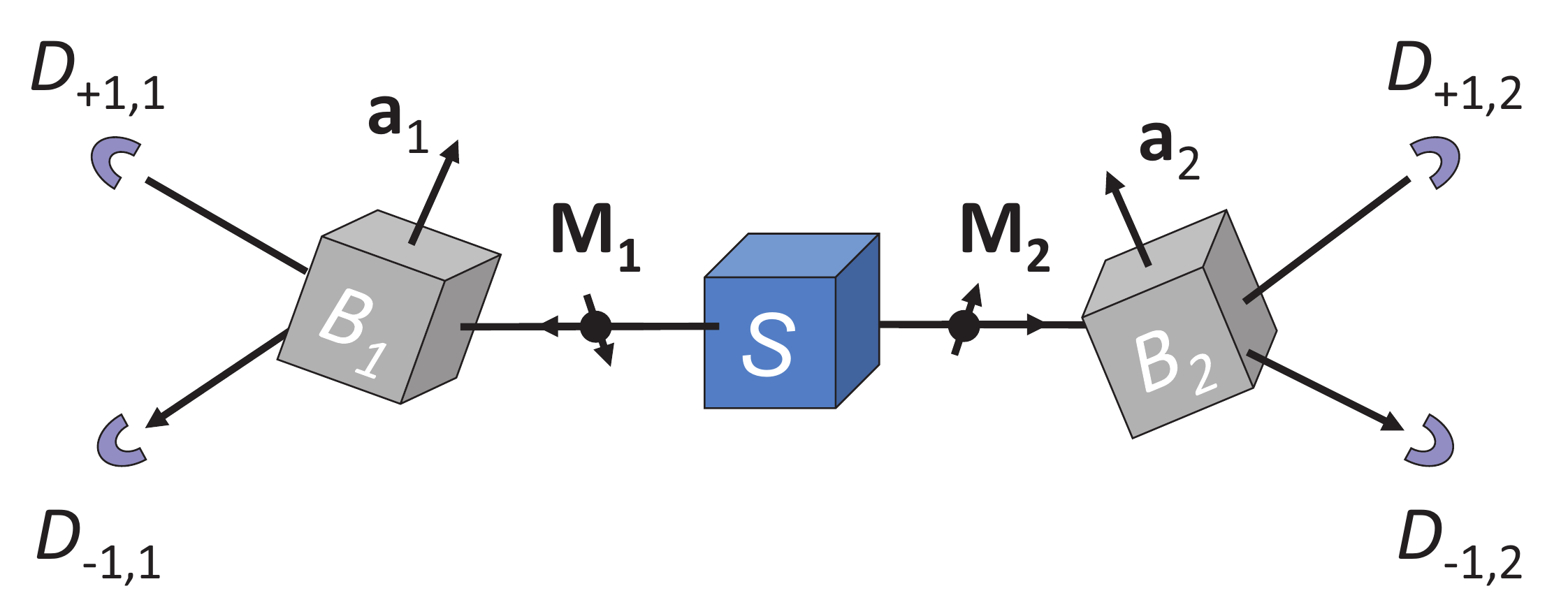}
\caption{(Color online)
Diagram of the EPRB thought experiment.
The source $S$, activated at times labeled by $i=1,2,\ldots,N$,
sends a particle with magnetic moment, represented by the unit vector $\mathbf{M_1}$,
to the Stern-Gerlach magnet $B_1$ and another particle
with magnetic moment, represented by the unit vector $\mathbf{M_2}$, to the Stern-Gerlach magnet $B_2$.
The orientations of the magnets, represented by unit vectors $\mathbf{a_1}$ and $\mathbf{a_2}$,
affect the deflection of the particles by the magnets.
Each particle going to the left (right) is detected with 100\% certainty by either
$D_{+1,1}$ or $D_{-1,1}$ ($D_{+1,2}$ or $D_{-1,2}$).
}
\label{EPRBthought}
\end{figure}

Referring to the schematic shown in Fig.~\ref{EPRBthought},
the i-prob to observe a pair $\{x,y\}$ is denoted by $P(x,y|\mathbf{a}_1,\mathbf{a}_2,Z)$
where $Z$ represents all the conditions
under which the experiment is performed, with exception of
the directions $\mathbf{a}_1$ and $\mathbf{a}_2$ of the Stern-Gerlach magnets $B_1$ and $B_2$, respectively.
It is important to note that $P(x,y|\mathbf{a}_1,\mathbf{a}_2,Z)$ does not depend on $\mathbf{M}_1$ and $\mathbf{M}_2$.
In concert with the general assumption (d), it is assumed that there is no relation between the actual values
of the pairs $\{x_i,y_i\}$ and $\{x_{i^\prime},y_{i^\prime}\}$ if $i\not=i^\prime$,
meaning that each repetition of the experiment represents an identical event of which
the outcome is logically independent of any other such event.
Invoking the product rule, the logical consequence of this assumption is that
$
P(x_1,y_1,\ldots,x_N,y_N|\mathbf{a}_1,\mathbf{a}_2,Z)=\prod_{i=1}^{N}P(x_i,y_i|\mathbf{a}_1,\mathbf{a}_2,Z)
$
meaning that the i-prob $P(x_1,y_1,\ldots,x_N,y_N|\mathbf{a}_1,\mathbf{a}_2,Z)$
to observe the compound event $\left\{\{x_1,y_1\},\ldots,\{x_N,y_N\}\right\}$
is completely determined by the i-prob $P(x,y|\mathbf{a}_1,\mathbf{a}_2,Z)$ to observe the pair $\{x,y\}$.

We also assume that the i-prob $P(x,y|\mathbf{a}_1,\mathbf{a}_2,Z)$
to observe a pair $\{x,y\}$ does not change if we apply
the same rotation to both magnets $B_1$ and $B_2$.
Expressing this invariance with respect to rotations
of the coordinate system (Euclidean space and Cartesian coordinates are used throughout this paper)
in terms of i-probs yields
$P(x,y|\mathbf{a}_1,\mathbf{a}_2,Z)=P(x,y|{\cal R}\mathbf{a}_1,{\cal R}\mathbf{a}_2,Z)$
where ${\cal R}$ denotes an arbitrary rotation in three-dimensional space which is
applied to both magnets $B_1$ and $B_2$, implying that $P(x,y|\mathbf{a}_1,\mathbf{a}_2,Z)$
is a function of the inner product $\mathbf{a}_1\cdot\mathbf{a}_2$ only.
Therefore, we must have
$P(x,y|\mathbf{a}_1,\mathbf{a}_2,Z)=P(x,y|\mathbf{a}_1\cdot\mathbf{a}_2,Z)=P(x,y|\theta,Z)$
where $\theta=\arccos(\mathbf{a}_1\cdot\mathbf{a}_2)$ denotes the angle between the unit vectors
$\mathbf{a}_1$ and $\mathbf{a}_2$.
Note that for any integer value of $K$, $\theta+2\pi K$
represents the same physical arrangement of the magnets $M_1$ and $M_2$.
From the algebra of LI, it follows that the i-prob to observe $x$, irrespective of the observed value of $y$ is given by
$P(x|\mathbf{a}_1,\mathbf{a}_2,Z)=\sum_{y=\pm1}P(x,y|\mathbf{a}_1,\mathbf{a}_2,Z)$.
In the context of the EPRB experiment, it is assumed that observing $x=+1$ is as likely as
observing $x=-1$, independent of the observed value of $y$.
This implies that we must have $P(x=+1|\mathbf{a}_1,\mathbf{a}_2,Z)=P(x=-1|\mathbf{a}_1,\mathbf{a}_2,Z)$ which,
in view of the fact that $P(x=+1|\mathbf{a}_1,\mathbf{a}_2,Z)+P(x=-1|\mathbf{a}_1,\mathbf{a}_2,Z)=1$
implies that $P(x=+1|\mathbf{a}_1,\mathbf{a}_2,Z)=P(x=-1|\mathbf{a}_1,\mathbf{a}_2,Z)=1/2$.
Applying the same reasoning to the assumption that, independent of the observed values of $x$,
observing $y=+1$ is as likely as observing $y=-1$
yields $P(y|\mathbf{a}_1,\mathbf{a}_2,Z)=P(x=+1,y|\mathbf{a}_1,\mathbf{a}_2,Z)+P(x=-1,y|\mathbf{a}_1,\mathbf{a}_2,Z)=1/2$.
Note that we did not assign any prior i-prob nor did we make any reference to concepts such as the singlet-state.
Although the symmetry properties which have been assumed are reminiscent of those of the singlet state,
this is deceptive: without altering the assumptions,
the LI approach yields the correlations for the triplet states as well~\cite{RAED14b}.

Adopting the same reasoning as in section~2b,
it follows directly from assumptions (a)--(d) that the i-prob to observe a pair $\{x,y\}$
takes the form~\cite{RAED14b}
$P(x,y|\theta,Z)=[1+xyE_{12}(\theta)]/4$
where
$E_{12}(\theta)=E_{12}(\mathbf{a}_1,\mathbf{a}_2,Z)$ is a periodic function of $\theta$.
Minimization of the corresponding expression of $|\mathrm{Ev}|$ while taking into account the
constraints (C2) and (C3) (see section~2a)
is tantamount to finding the i-prob's $P(x,y|\theta,Z)$ which minimize~\cite{RAED14b}
\begin{equation}
I_{F}= \sum_{x,y=\pm1}
\frac{1}{P(x,y|\theta,Z)}
\left(\frac{\partial P(x,y|\theta,Z)}{\partial\theta}\right)^2
=\frac{1}{1-E_{12}^2(\theta)}
\left(\frac{\partial E_{12}(\theta)}{\partial \theta}\right)^2,
\label{epr6}
\end{equation}
subject to the constraint that $\partial P(x,y|\theta,Z)/\partial\theta\not=0$
for some pairs $(x,y)$.
Equation.~(\ref{epr6}) is readily integrated to yield
$E_{12}(\theta)=\cos\left(\theta\sqrt{I_F}+\phi\right)$
where $\phi$ is an integration constant.
As $E_{12}(\theta)$ is a periodic function of $\theta$
we must have $\sqrt{I_F}=K$ where $K$ is an integer and hence
$E_{12}(\theta)=\cos\left(K\theta+\phi\right)$.
Because of constraint (C2)
we exclude the case $K=I_F=0$ from further consideration.
Hence the physically relevant, nontrivial solution corresponds to $K=1$.
Furthermore, as $E_{12}(\theta)$ is a function of
$\mathbf{a}_1\cdot\mathbf{a}_2=\cos\theta$ only,
we must have $\phi=0,\pi$, reflecting an ambiguity
in the definition of the direction of $B_1$ relative to the direction of $B_2$.
Choosing the solution with $\phi=\pi$, we find
\begin{eqnarray}
P(x,y|\mathbf{a}_1,\mathbf{a}_2,Z)&=&\frac{1-xy\mathbf{a}_1\cdot\mathbf{a}_2}{4}
\quad,\quad
\langle xy\rangle =\sum_{x,y=\pm1} xyP(x,y|\mathbf{a}_1,\mathbf{a}_2,Z)=
-\mathbf{a}_1\cdot\mathbf{a}_2
,
\label{P2D9}
\end{eqnarray}
$\langle x\rangle = \sum_{x,y=\pm1} xP(x,y|\mathbf{a}_1,\mathbf{a}_2,Z) =0$,
and $\langle y\rangle = \sum_{x,y=\pm1} yP(x,y|\mathbf{a}_1,\mathbf{a}_2,Z) =0$,
all in complete agreement with the quantum theoretical description of two $S=1/2$
particles in the singlet state~\cite{BOHM51,BALL03}.
As the LI treatment of a robust EPPB experiment directly yields the probabilistic
description that we know from quantum theory without invoking the notions the latter,
it follows that the concept of quantum entanglement cannot be essential
for describing the data produced by EPRB experiments.

It may be of interest to mention here that in spite of the widely spread claims
that real EPRB experiments have proven quantum theory correct, none of the three different
experiments for which data has been made available~\cite{WEIH00,SHIH11,VIST12}
survives the confrontation with the 5-standard-deviation-criterion
hypothesis test that the data complies with the quantum theoretical description
given by Eq.~(\ref{P2D9})~\cite{RAED13a,HESS15}.
It seems that for the time being, only computer experiments are able to
generate data that are not in conflict with the quantum theoretical description
of the EPRB thought experiment~\cite{RAED12a,RAED13a}.

\subsection{Application: Particle in a potential}\label{sec2c}

For simplicity of notation and presentation, in the present review, we only discuss the problem
of inferring the plausibility that the particle is at a certain position $X$ on a line
and produces a click on a detector at position $x$, also on a line.
The full fledged three-dimensional derivation including electromagnetic
potentials and/or spin can be found elsewhere~\cite{RAED14b,RAED15b}.

The measurement scenario is as follows.
We imagine $N$ repetitions ($n=1,\ldots,N$) of an experiment performed on a particle moving on a line of linear extent $[-L,L]$.
Nothing is known about the direction of motion of the particle.
In each such experiment, a source emits a signal at discrete times labeled by the integer $\tau=1,\ldots,M$.
It is assumed that for each repetition, the particle is at the unknown position $-L \le X_\tau\le L$.
the signal solicits a response of the particle, generating a click of detector at discrete position
$j_{n,\tau}$ with $-K\le j_{n,\tau}\le K$.
The detectors $-K,\ldots,K$ have spatial extent $\Delta=L/K$
and are placed next to each other, completely covering the line segment $[-L,L]$.
It is assumed that for each signal emitted by the source, one and only one of the $2K+1$ detectors fires.

The result of $N$ repetitions of the experiment yields the data set
\begin{equation}
\Upsilon=\{j_{n,\tau}| -K \le j_{n,\tau}\le K\;;\; n=1,\ldots,N\;;\; \tau=1,\ldots,M \}
,
\label{TDSE0}
\end{equation}
or, denoting the total count of clicks of detector $j$ at time $\tau$ by $0\le k_{j,\tau}\le N$, we have
\begin{eqnarray}
{\cal D}&=&\Bigl\{ k_{j,\tau}\Bigl|  N=\sum_{j=-K}^{K} k_{j,\tau}\;;\; \tau=1,\ldots,M \Bigr\}
.
\label{TDSE1}
\end{eqnarray}
Following the general procedure, the next step is to
introduce the i-prob $P(j|X_\tau,\tau,Z)$, expressing the relation between
a particle at unknown location $X_\tau$ at discrete time $\tau$ and the click of the detector at position $j$.
The conditions represented by $Z$ are fixed and identical for all experiments.
Note that we will not try to estimate the unknown position $X$ but
rather determine the i-prob $P(j|X_\tau,\tau,Z)$ which yields the most robust set of data,
robust with respect to small changes of $X_\tau$ for all $\tau$.
According to basic assumption (d), there is no relation between the actual values of $j_{n,\tau}$ and $j_{n',\tau'}$
if $n\not=n'$ or $\tau\not=\tau'$.
Hence, for fixed positions $X_\tau$, the i-prob to observe all the data ${\cal D}$ is given by
\begin{equation}
P({\cal D}|X_1,\ldots,X_N,N,Z) = N!\prod_{\tau=1}^M\prod_{j=-K}^K \frac{P(j|X_\tau,\tau,Z)^{k_{j,\tau}}}{k_{j,\tau}!}
.
\label{TDSE2}
\end{equation}
It is now straightforward to repeat the steps that led to Eq.~(\ref{robu1})
to find that the measure of a robust experiment is given by
\begin{equation}
\mathrm{Ev}=
-\frac{N\epsilon^2}{2}
\sum_{j=-K}^K\;\;
\sum_{\tau=1}^M
\frac{1}{P(j|X_\tau,\tau,Z)}
\left(\frac{\partial P(j|X_\tau,\tau,Z)}{\partial X_\tau}\right)^2
+{\cal O}(\epsilon^3)
,\label{SE5}
\end{equation}
and that the most nontrivial robust experiment is described by
the i-prob $P(j|X_\tau,\tau,Z)$ which minimizes the Fisher information
\begin{equation}
I_F=
\sum_{j=-K}^K\;\;
\sum_{\tau=1}^M
\frac{1}{P(j|X_\tau,\tau,Z)}
\left(\frac{\partial P(j|X_\tau,\tau,Z)}{\partial X_\tau}
\right)^2
,
\label{TDSE8}
\end{equation}
subject to the constraint that not all $\partial P(j|X_\tau,\tau,Z)/\partial X_\tau$ are zero
and the additional constraints to be discussed below.

As the Schr\"odinger equation is formulated in continuum space,
it is necessary to replace Eq.~(\ref{TDSE8}) by its continuum limit
\begin{equation}
I_F=
\int \; dx \int\; dt \;
\frac{1}{P(x|X(t),t,Z)}
\left(\frac{\partial P(x|X(t),t,Z)}{\partial X(t)}
\right)^2
=
\int \; dx \int\; dt \;
\frac{1}{P(x|X(t),t,Z)}
\left(\frac{\partial P(x|X(t),t,Z)}{\partial x}
\right)^2
,
\label{SE6a}
\end{equation}
where we assumed that it does not matter where in space we perform the experiment
(homogeneity of space), implying that $P(x|X(\tau),\tau,Z)=P(x+\zeta|X(\tau)+\zeta,\tau,Z)$
where $\zeta$ is an arbitrary real number.
As before, it is a symmetry requirement which allows us to regard
the unknown quantity $X$ as the ``coordinate of a particle''
based on measurement of the coordinate of the detector that clicks.
Technically speaking, after passing to the continuum limit, $P(x|X(t),t,Z)$
denotes the probability density, not the probability itself but as there can be no confusion
about which case, discrete or continuum, we are considering,
we use the same symbol for the probability density and the probability.

In general, if there is no uncertainty about individual events, we expect the description to
agree with classical theoretical mechanics.
We use this ``correspondence principle'' to incorporate
classical theoretical mechanics into the LI approach~\cite{RAED14b,RAED15b}.
In the absence of uncertainty and in line with the basic ideas of classical mechanics,
the observed detector clicks form smooth trajectories.
One such trajectory can always be represented by $d x(t)/dt = U(x(t),t)$
but the function $U(.,.)$ may not be ``universal'' in the sense that it may
change from experiment to experiment, e.g. with the initial conditions.
However, if $U(.,.)$ is universal and sufficiently ``nice'' (we ignore technical details related
to differentiability etc.) we may write~\cite{RALS13b}
\begin{equation}
\frac{d x(t)}{dt}= U(x(t),t) = \frac{\partial S(x,t)}{\partial x}
,
\label{MO0}
\end{equation}
where $S(x,t)\equiv S(x(t),t)$ and it follows that
\begin{equation}
\frac{d^2 x(t)}{dt^2}= \frac{\partial }{\partial x} \left[
\frac{\partial S(x,t)}{\partial t}+
\frac{1}{2}\left(\frac{\partial S(x,t)}{\partial x}\right)^2
\right]\equiv - \frac{\partial V(x,t)}{\partial x}
,
\label{MO1}
\end{equation}
showing that if there exists a universal function $U(x(t),t)$ which describes the data
according to Eq.~(\ref{MO0}), then there exists a potential $V(x,t)$ such that the Hamilton-Jacobi equation
\begin{equation}
\frac{\partial S(x,t)}{\partial t}+ \frac{1}{2}\left(\frac{\partial S(x,t)}{\partial x}\right)^2 + V(x,t) =0
,
\label{MO2}
\end{equation}
holds~\cite{RALS13b}.
Thus, the assumption that in the absence of uncertainty, all the possible trajectories $x(t)$
can be described by one function $U(.,.)$ quite straightforwardly yields Eq.~(\ref{MO2}),
that is one of the formulations of classical theoretical mechanics.

In the presence of uncertainty about individual events, we can combine the notion of a robust experiment
and the desire to recover equations of classical mechanics as a limiting case
by searching for the robust (i.e. for all $X(t)$) minima of the functional~\cite{RAED14b,RAED15b}
\begin{eqnarray}
F&=&
\int dx\; \int dt\;
\bigg\{
\frac{1}{P(x|X(t),t,Z)}
\left(\frac{\partial P(x|X(t),t,Z)}{\partial x}\right)^2
\nonumber \\
&&\hbox to 2cm{}+2m\lambda
\left[
\frac{\partial S(x,t)}{\partial t}+\frac{1}{2m}\left(\frac{\partial S(x,t)}{\partial x}\right)^2+V(x,t)
\right]
P(x|X(t),t,Z)
\bigg\}
,
\label{TDSE10}
\end{eqnarray}
where $\lambda$ is a parameter having dimension $\mathrm{s}^2/\mathrm{kg}\; \mathrm{m}^4$ 
and, for convenience of comparing with quantum theory, we introduced the mass $m$ of the particle
by substituting $S(x,t)\rightarrow S(x,t)/m$ and $V(x,t)\rightarrow V(x,t)/m$.

Standard variational calculus yields the extrema of Eq.~(\ref{TDSE10}) in terms of two coupled non-linear first-order
differential equations of the functions $P(x|X(t),t,Z)$ and $S(x,t)$ which are identical to the (one-dimensional version of the)
equations that appear in Madelung's hydrodynamical form~\cite{MADE27} or Bohm's interpretation~\cite{BOHM52} of quantum theory.
However, Eq.~(\ref{TDSE10}) was not derived from quantum theory but was obtained through logic inference
from data produced by robust experiments and a correspondence principle, without invoking concepts of quantum theory.
Therefore, in principle we do not need the latter to describe
these experiments but we can use the equivalence of Eq.~(\ref{TDSE10}) and the mathematical framework of quantum
theory to great advantage for turning the non-linear equations into linear ones which can be solved
by the powerful machinery of linear algebra.
Technical details of the derivation of the functionals analogous to $F$ for the
multidimensional Schr\"odinger equation and
Pauli equation for a particle with spin can be found in the original papers~\cite{RAED14b} and~\cite{RAED15b}, respectively.

\section{Connecting with quantum theory}\label{sec3}

The LI approach yields descriptions of robust experiments in terms of i-probs.
In this section, we discuss two different methods
of transforming these i-probs into the wavefunction-formalism of QT.

The first method is based on the general observation that in
scientific reasoning it is good practice to reduce the complexity of the description of the whole
by separating the description of data into several parts.
We consider different ways of organizing the observed data
and scrutinize the conditions under which a description of the various parts
of the experiment can be separated (as much as possible).
Then we show, in the case of the Stern-Gerlach and EPRB experiment, how
the wavefunction description naturally emerges as a result of this separation procedure.
It automatically follows that the wavefunction (or density matrix) description
is less general than the one in terms of conditional probabilities in the sense that
the former can only describe situations in which the separation procedure can actually be carried out.

The second method employs a polar representation of the i-prob to
bring the non-linear robust optimization problem into a linear form and is
most useful for cases that involve dynamics, yielding Schr\"odinger-like equations.

\subsection{Separation procedure~\cite{RAED15c}}\label{sec3a}

Consider again the Stern-Gerlach experiment, see Fig.~\ref{SGthought}, yielding the data set
\begin{eqnarray}
{\cal D} &=& \big\{x_1,\ldots,x_N|x_i=\pm1\;,i=1,\ldots,N\big\}
,
\label{SG0}
\end{eqnarray}
where $N$ is total the number of recorded events.
Suppose that the analysis of correlations among the observed $x_i$
indicates that the $x_i$ are independent events, in line with assumption (d) (see Section~\ref{sec2}).
Then the counts $N(\pm1|\mathbf{a},\mathbf{M},Z)$ of outcomes with
$x=\pm1$ ($N=N(+1|\mathbf{a},\mathbf{M},Z)+N(-1|\mathbf{a},\mathbf{M},Z)$)
give a complete characterization of the data.
In essence, all data sets having the same average
\begin{eqnarray}
\langle x \rangle &=& \frac{1}{N}\sum_{x=\pm1} x N(x|\mathbf{a},\mathbf{M},Z) \equiv \sum_{x=\pm1} x f(x|\mathbf{a},\mathbf{M},Z)
,
\label{SG1}
\end{eqnarray}
are equivalent.
Assuming (as in Section~\ref{sec2}a) that the observed counts do not depend on the orientation
of the chosen reference frame, $f(x|\mathbf{a},\mathbf{M},Z)$ can only depend on $\mathbf{a}\cdot\mathbf{M}$
(by construction $\vert\mathbf{a}\vert=1$ and $\vert\mathbf{M}\vert=1$).
Hence, we must have $f(x|\mathbf{a},\mathbf{M},Z)=f(x|\mathbf{a}\cdot\mathbf{M},Z)$.

Equation.~(\ref{SG1}) is a holistic description of the data in terms of $\mathbf{a}\cdot\mathbf{M}$
and it is by no means obvious how to construct, if possible at all, a description
in terms of a part that refers to the object (represented by $\mathbf{M}$)
and another part that refers to the magnet (represented by $\mathbf{a}$).
To explore the possibilities of separating in parts, it is expedient to consider
alternative ways of writing Eq.~(\ref{SG1}).
Let us first organize the data and frequencies in vectors $\mathbf{x}=(+1,-1)^T$
and $\mathbf{f}=(f(+1|\mathbf{a},\mathbf{M},Z),f(-1|\mathbf{a},\mathbf{M},Z))^T$, respectively.
Then, we trivially have
\begin{equation}
\langle x \rangle = \mathbf{x}^T\cdot\mathbf{f}= \mathbf{Tr\;} \mathbf{x}^T\mathbf{f}= \mathbf{Tr\;}\mathbf{f} \mathbf{x}^T
,
\label{P2D0a}
\end{equation}
where $\mathbf{f} \mathbf{x}^T$ is a $2\times2$ matrix and $\mathbf{Tr}\; A $ denotes the trace of the matrix $A$.
Now note that {\bf any} rewriting of $\mathbf{x}$
and $\mathbf{f}$ in terms of vectors, matrices, \ldots,
$\widetilde{\mathbf{x}}$ and $\widetilde{\mathbf{f}}$ such that
$\mathbf{Tr\;}\widetilde{\mathbf{f}}=1$ and
$\mathbf{Tr\;}\widetilde{\mathbf{f}}\widetilde{\mathbf{x}}=\langle x \rangle$
does not change $\langle x \rangle$, that is it yields the same complete description of the data.
Therefore, with this in mind, we consider the rearrangement of the data
into $2\times2$ (diagonal, hermitian) matrices $X$ and $F$ with elements
$X(x,x')=x\delta_{x,x'}$ and
$F(x,x')=f(x|\mathbf{a},\mathbf{M},Z)\delta_{x,x'}$, respectively
and rewrite Eq.~(\ref{P2D0a}) as
\begin{eqnarray}
\langle x \rangle &=& \mathbf{Tr}\; FX = \mathbf{Tr}\; \widehat\rho \widehat X
,
\label{P2D0}
\end{eqnarray}
where $\widehat\rho$ and $\widehat X$ can be any pair of $2\times2$ matrices that satisfies Eq.~(\ref{P2D0}).
Clearly, a formal rewriting of Eq.~(\ref{P2D0a}) such as Eq.~(\ref{P2D0}) cannot, by itself, bring anything new
but representation Eq.~(\ref{P2D0}) offers the flexibility that allows us to perform the separation
by using some elementary linear algebra, as we now show.

We know from linear algebra that any hermitian $2\times2$ matrix
can be written as a linear combination of four hermitian $2\times2$ matrices.
Without loss of generality, we may choose the Pauli-spin matrices $\bm{\sigma}=(\sigma^x,\sigma^y,\sigma^z)$
and the unit matrix $\openone$ as the orthonormal basis set
for the vector space of $2\times2$ matrices with an inner product defined by $(A,B)=\mathbf{Tr}\; A^\dagger B$.
Without loss of generality we may write
\begin{equation}
\widehat\rho=\frac{\openone+\bm\rho\cdot\bm{\sigma}}{2}
\quad,\quad
\widehat X=u_0\openone+\mathbf{u}\cdot\bm{\sigma}
.
\label{P2D1}
\end{equation}
where  $\bm\rho=(\rho_x,\rho_y,\rho_z)$, $u_0$, and $\mathbf{u}=(u_x,u_y,u_z)$
are all real-valued. It is now straightforward to show~\cite{RAED15b} that
the desired separation  can be realized by requiring that
$u_0=u_0(\mathbf{a},Z)$,
$u_x=u_x(\mathbf{a},Z)$,
$u_y=u_y(\mathbf{a},Z)$,
$u_z=u_z(\mathbf{a},Z)$,
$\rho_x=\rho_x(\mathbf{M},Z)$,
$\rho_y=\rho_y(\mathbf{M},Z)$,
and
$\rho_z=\rho_z(\mathbf{M},Z)$ (recall that $Z$ is considered to represent all fixed conditions
which are important to the actual experiment but are not of immediate interest).
Assuming that the observed counts do not depend on the orientation of the reference frame (see earlier),
$\langle x \rangle$ is a function of $\mathbf{a}\cdot\mathbf{M}$ only.
This requirement enforces $\bm{\rho}=\mathbf{M}$, and $\mathbf{u}=\mathbf{a}$.
Hence, we have $\langle x \rangle=u_0+\mathbf{M}\cdot\mathbf{a}$ and
as $|\langle x \rangle|\le 1$ it follows that $-1\le u_0+\mathbf{M}\cdot\mathbf{a}\le1$.
for $\mathbf{a}=\mathbf{M}$ and $\mathbf{a}=-\mathbf{M}$ we have $u_0\le 0 $ and $0\ge u_0$, respectively, hence $u_0=0$.
Note that we could equally well have made the choice $\bm{\rho}=\mathbf{a}$, and $\mathbf{u}=\mathbf{M}$
instead of $\bm{\rho}=\mathbf{M}$, and $\mathbf{u}=\mathbf{a}$.
However, the former choice leads to inconsistencies
for instance when we consider an experiment in which we place several SG magnets
in succession or consider the EPRB experiment.

Thus, we have shown that only the desire
to represent the data Eq.~(\ref{SG0}) such that the description of the whole experiment
separates into a description of the ``source'' ($\mathbf{M}$)
and a description of the ``measurement device'' ($\mathbf{a}$)
together with some elementary linear algebra leads to the unique
description of the SG experiment in terms of $2\times2$ matrices
\begin{equation}
\widehat\rho=(\openone+\mathbf{M}\cdot\bm{\sigma})/2
\quad,\quad
\widehat X=\mathbf{a}\cdot\bm{\sigma}
,
\label{P2D6}
\end{equation}
conditional on the assumptions that the individual outcomes of a SG experiment
are independent and that the frequency distribution of these outcomes
does not depend on the orientation of the reference frame.
From Eq.~(\ref{P2D6}) it follows immediately that $\widehat\rho^2=\widehat\rho$, that is $\widehat\rho$ is a projection.
This implies that we can write~\cite{BALL03}
\begin{equation}
\widehat\rho=|\Psi\rangle \langle\Psi|
\quad,\quad
|\Psi\rangle=a_\uparrow|\uparrow\rangle+a_\downarrow|\downarrow\rangle
,
\label{P2D6a}
\end{equation}
where the vector $|\Psi\rangle$ is expressed in the basis of the eigenstates ($|\uparrow\rangle$,$|\downarrow\rangle$)
of the $\sigma^z$ matrix.

Summarizing: changing the representation of the data in combination
with the desire to separate as much as possible the description of
the source and measurement devices automatically enforces the Hilbert space
structure that is a characteristic signature of quantum theory~\cite{RAED15b}.
No postulates of quantum theory are required to derive (or postulate) Eqs.~(\ref{P2D6}) or (\ref{P2D6a}).
Furthermore, it is straightforward to extent the description to include mixed states~\cite{RAED15b}.

As there is nothing that forbids an experiment to yield for instance
$f(x|\mathbf{a}\cdot\langle\mathbf{M}\rangle),Z)=(1+x(\mathbf{a}\cdot\mathbf{M})^k)/2$
with $k=2$ for instance (we certainly can generate such data using a digital computer, a metaphor
of a physical device on which we carry out experiments).
However, the data produced by such an experiment cannot be represented by Eq.~(\ref{P2D6}).
In other words, the class of conceivable SG experiments is significantly larger
than the class of experiments that allows for the separation:
the class of realizable SG experiments is (much) larger than the class of
SG experiments describable by quantum theory.

The fact that the separation procedures leads, in such simple manner,
to the quantum theoretical description Eq.~(\ref{P2D6}) of the SG experiment
provokes to question ``what is so special about the case in which the separation   procedure
can be carried out?''
The answer is given in Section~\ref{sec2}.
Using Eq.~(\ref{P2D6}), according to the postulate of quantum theory~\cite{BALL03},
the probability to observe an event $x$ is given by
$P(x)=\mathbf{Tr\;}\widehat\rho (\openone+x\mathbf{a}\cdot\bm{\sigma})/2 = (1+x \mathbf{a}\cdot\mathbf{M})/2$
which is exactly the same expression as the one obtained by LI treatment of a robust SG experiment.
In other words, if the SG experiment is robust, it may be equally well be described by quantum theory.

The application of the separation procedure to the EPRB experiment is an almost trivial extension
of the application to the SG experiment.
We start by writing the observations $xy=(+1,-1,+1,-1)$
and frequencies $(f(+1,+1|\theta,Z)$, $f(-1,+1|\theta,Z)$, $f(+1,-1|\theta,Z)$, $f(-1,-1|\theta,Z))$
as $4\times4$ matrices $X$ and $F$ with elements
$X([x,y],[x',y'])=x\delta_{x,x'}\delta_{y,y'}$ and
$Y([x,y],[x',y'])=y\delta_{x,x'}\delta_{y,y'}$ and
$F([x,y],[x',y'])=f(xy|\mathbf{a},\mathbf{M},Z)\delta_{x,x'}\delta_{y,y'}$, respectively.
Here we use the notation $[x,y]=(1-x)/2+(1-y)$ to indicate that the pairs $(x,y)$ and
$(x',y')$ specify the row, respectively the column index (running from 0 to 3)
of the matrices $X$ and $F$.
We search for $4\times4$ matrices $\widehat\rho$, $\widehat X$, and  $\widehat Y$
which satisfy
\begin{eqnarray}
\mathbf{Tr} \widehat\rho =1
\quad,\quad
\mathbf{Tr} \widehat\rho \widehat X=\langle x\rangle
\quad,\quad
\mathbf{Tr} \widehat\rho \widehat Y=\langle y\rangle
\quad,\quad
\mathbf{Tr} \widehat\rho \widehat X \widehat Y=\langle xy\rangle
,
\label{sec3a0}
\end{eqnarray}
and allow for the desired separation.
Using the direct product of the Pauli-spin matrices
$\bm{\sigma}_j=(\sigma^x_j,\sigma^y_j,\sigma^z_j)$ for $j=1,2$
and the unit matrix $\openone$ as the orthonormal basis set for the vector space of
$4\times4$ matrices, we may write (without loss of generality)
%
$\widehat\rho=\rho_0\openone+\bm\rho_1\cdot\bm{\sigma}_1\otimes\openone_2+\openone_1\otimes\bm\rho_2\cdot\bm{\sigma}_2
+\bm{\sigma}_1\cdot\bm\rho_{12}\cdot\bm{\sigma}_2
$.
where the number $\rho_0$, the vectors $\bm\rho_j$, and the matrix $\bm\rho_{12}$ are all real-valued.
As each of the two sides of the EPRB experiment contains a SG magnet,
consistency with the separated   description of the SG experiment demands that we choose
$
\widehat X=\mathbf{a}_1\cdot\bm{\sigma}_1\otimes\openone_2$ and
$Y=\openone_1\otimes\mathbf{a}_2\cdot\bm{\sigma}_2$.
We find the explicit expression of $\widehat\rho$ by requiring that Eq.~(\ref{sec3a0}) holds.
Focussing on the case of the EPRB experiment
for which $\langle x\rangle=\langle y\rangle=0$ and
$\langle xy\rangle=-\mathbf{a}_1\cdot\mathbf{a}_2$,
it follows that $\rho_0=1/4$, $\bm\rho_1=\bm\rho_2=0$
and that $\widehat\rho$ takes the form~\cite{RAED15b}
\begin{eqnarray}
\widehat\rho&=&\frac{\openone-\sigma_1\cdot\sigma_2}{4}
.
\label{P2D12}
\end{eqnarray}
It is not difficult to verify that ${\widehat\rho}^2={\widehat\rho}$, hence
Eq.~(\ref{P2D12}) is the density matrix of a pure quantum state~\cite{BALL03}.
Computing the matrix elements of ${\widehat\rho}$ in the
spin-up, spin-down basis of both spins we find
\begin{eqnarray}
\widehat\rho&=&
\left(\frac{|\uparrow\downarrow\rangle-|\downarrow\uparrow\rangle}{\sqrt{2}}\right)
\left(\frac{\langle\uparrow\downarrow|-\langle\downarrow\uparrow|}{\sqrt{2}}\right)
,
\label{P2D12a}
\end{eqnarray}
and
\begin{eqnarray}
\langle xy\rangle&=&\mathbf{Tr\;} \widehat\rho \widehat X \widehat Y
=\mathbf{Tr\;} \widehat\rho \mathbf{a}_1\cdot\bm{\sigma}_1\mathbf{a}_2\cdot\bm{\sigma}_2
=
-\mathbf{a}_1\cdot\mathbf{a}_2
,
\label{P2D12b}
\end{eqnarray}
which we recognize as the quantum theoretical description of two spin-1/2 objects in the singlet state.
Therefore, we have shown that rewriting the data gathered in an ideal EPRB thought experiment in a manner
that allows for the envisaged separation  naturally leads, without invoking postulates of quantum theory
and/or probability theory, to the quantum theoretical description of two $S=1/2$ spins in the singlet state.

As in the case of the ideal SG experiment, the representation in parts
puts a severe restriction on the kind of data that we can describe, again
provoking to question ``what is so special about the case in which the separation   procedure
can be carried out?''
The answer is the same as in the case of the SG experiment:
it is precisely for the special case of the robust EPRB experiment.


\subsection{Equivalence with quadratic forms}\label{sec3c}

In the case of SG or EPRB experiments, the LI approach yields equations for the i-probs which are easy to solve directly.
As the i-probs describe the data produced by robust experiments, the connection to the quantum formalism is mainly of pedagogical interest.
However, if the equations for the i-probs are non-linear, as in the case of a particle in a potential discussed in Section~2c,
and not easy to solve, it is expedient to search for alternative equations that are much easier to solve.
Fortunately, in the case at hand, we can make good use of the large body of work that explores mathematically equivalent forms of quantum theory.

Consider the quadratic functional
\begin{equation}
Q=\int dx\; \int dt\;
\bigg[
2im\sqrt{\lambda}
\left(
\psi\frac{\partial \psi^\ast}{\partial t} - \psi^\ast\frac{\partial \psi}{\partial t}
\right)
+
4\frac{\partial \psi^\ast}{\partial x}
\frac{\partial \psi}{\partial x}
+2m\lambda V(x,t)\psi^\ast\psi
\bigg]
,
\label{TDSE11}
\end{equation}
with the shorthand notation $\psi\equiv\psi(x|X(t),t,Z)$.
Substitute $\psi=\sqrt{P(x|X(t),t,Z)}e^{iS(x,t)\sqrt{\lambda}/2}$ to obtain Eq.~(\ref{TDSE10}),
demonstrating that Eq.~(\ref{TDSE10}) and Eq.~(\ref{TDSE11}) are equivalent (the ambiguity in
the phase of $\psi$ can be shown to be irrelevant~\cite{RAED14b}).
On the other hand, the extrema of Eq.~(\ref{TDSE11}) are given by the solution of
the linear partial differential equation
\begin{equation}
\frac{2i}{\sqrt{\lambda}}\frac{\partial \psi}{\partial t}
=
-\frac{2}{m\lambda}\frac{\partial^2 \psi}{\partial x^2} + V(x,t)\psi
,
\label{TDSE9a}
\end{equation}
which turns into the time-dependent Schr\"odinger equation if we set $\lambda=4/\hbar^2$.

From our derivation of Eq.~(\ref{TDSE9a}) from LI principles,
it is clear that (i) the actual value of $\lambda$ can only be determined by comparing the outcome
of calculations based on Eq.~(\ref{TDSE10}) or Eq.~(\ref{TDSE9a}) with experimental data
and that (ii) the wavefunction $\psi(x|X(t),t,Z)$ is just a mathematical concept,
a vehicle to solve a class of complicated nonlinear minimization problems
through the minimization of quadratic forms.
As a product of human imagination, this concept is an extraordinarily useful
tool that serves no purpose other than transforming nonlinear equations into linear ones.

\section{Conclusion}\label{sec4}

Using the simplest, non-trivial examples, it was shown how the application of LI
to experiments for which the observed events are independent
and for which the frequency distribution of these events is robust with respect to small changes of
the conditions under which experiments are carried out yields,
without introducing any concept of quantum theory, some of the most basic equations of quantum theory.
More extensive discussions of applications to the time-(in)dependent phenomena
with or without spin can be found elsewhere~\cite{RAED13b,RAED14a,RAED14b}.
Work to include relativistic effects is in progress~\cite{DONK15}.

The key point of a LI application to quantum physics experiments is to express
precisely and unambiguously, using the mathematical
framework of plausible reasoning~\cite{COX46,COX61,TRIB69,SMIT89,JAYN03}, the
conditions of a robust experiment, see Section~\ref{sec2}.
This translates into a global optimization problem for the i-prob,
the solution of which may be very simple as in the case of the SG and EPRB
experiment or may yield a fairly complicated non-linear set of equations.
The mathematical machinery of quantum theory appears as a result of transforming a set of non-linear equations
into a set of linear ones or emerges from the desire to separate the description into various parts.

It will not have escaped the reader that the LI approach reviewed in the present paper
is void of postulates regarding ``wavefunctions'', ``observables'', ``quantization rules'',
``quantum measurements''~\cite{NIEU13},``Born's rule'', etc. nor that there are ``interpretational'' issues.
This is a direct consequence of the basic premise of the LI approach, namely that
current scientific knowledge derives, through cognitive processes in the human brain, from the discrete events
which are observed in laboratory experiments and from relations between those events that we, humans, discover.
These discrete events are not ``generated'' according to certain quantum laws: instead these laws
appear as the result of (the best) LI from the data.

This viewpoint seems completely in line with Bohr's view~\cite{BOHR99}:
``Physics is to be regarded not so much as the study of something a priori given,
but rather as the development of methods of ordering and surveying human experience.
In this respect our task must be to account for such experience in a manner independent
of individual subjective judgment and therefore objective in the sense that it can be
unambiguously communicated in ordinary human language.''
This, in our opinion, is exactly what the LI approach allows us to do.
The extraordinary descriptive power of quantum theory then follows from the fact
that it is plausible reasoning, that is common sense, applied to robust experiments.

From our LI derivations of some of the most basic equations of quantum theory
it follows that the latter describes only robust experiments.
This is best illustrated by comparing the (high) accuracy by which quantum theory
predicts say, the ratios of the wavelengths of the Balmer absorption/emission lines of hydrogen~\cite{CODA12}
with the comparably low accuracy of say EPRB experiments that purport to provide evidence
for the singlet state of two spin-1/2 particles~\cite{RAED12a,RAED13a}.
In the former case, the high accuracy originates from doing a massive amount of experiments
on a very large collection of identical atoms and, as in any statistical experiment,
what we observe most of the time is the most robust response.
Thus, the solution of the LI problem (e.g. the Schr\"odinger equation
in the case at hand) is the one that is ``observed'' most frequently.
In contrast, in experiments that provide data on an event-by-event basis,
the statistical samples are much smaller and the external conditions may vary significantly
from one experiment to the next.
In other words, these experiments are not as robust as the spectroscopic experiments.
From the LI viewpoint it is therefore natural that these experiments produce data
that show (much) larger deviations from the quantum-theoretical prediction than spectroscopic data.
By the same argument, the LI approach offers a rational explanation for the observation
that it seems to take considerable effort to engineer nanoscale devices that operate in a regime
such that the experimental data complies with quantum theory.

%
%


\funding{MIK acknowledges financial support by European Research Council, project 338957 FEMTO/NANO.}

\conflict{The authors have no competing interests.}

\contributions{
All authors contributed to the material presented in and writing of this manuscript.
}



\end{document}